\pdfoutput=1
\documentclass[%
 reprint,
 amsmath,amssymb,
 aps,
]{revtex4-1}

\usepackage{graphicx}
\usepackage{dcolumn}
\usepackage{bm}

\usepackage{amsmath}	
\usepackage{amssymb}	
\newcommand{\msun}{{\rm M}_\odot}
\newcommand{\zsun}{{\rm Z}_\odot}
\usepackage{newtxtext,newtxmath}

\usepackage[normalem]{ulem}
\usepackage{physics}

\usepackage{xcolor}

\usepackage{newtxtext,newtxmath}
\usepackage[normalem]{ulem}

\usepackage{aas_macros}

\begin{document}

\title{The formation of black holes from rapidly accreting supermassive stars is not trivial: \\ Simulations of thermonuclear pulsations and explosions}

\author{Chris Nagele}
 \email{chrisnagele.astro@gmail.com}
\affiliation{Department of Astronomy, School of Science, The University of Tokyo, 7-3-1 Hongo, Bunkyo, Tokyo 113-0033, Japan}
\affiliation{William H. Miller Department of Physics and Astronomy, Johns Hopkins University, Baltimore, MD 21218, USA}
\author{Hideyuki Umeda}
\affiliation{Department of Astronomy, School of Science, The University of Tokyo, 7-3-1 Hongo, Bunkyo, Tokyo 113-0033, Japan}

\date{\today}

\begin{abstract}
In recent years, the formation and evolution of rapidly accreting supermassive stars has received significant attention in the hope of better understanding the origin of high redshift quasars. It is often taken for granted that once formed, these supermassive stars will encounter the general relativistic radial instability and collapse to form massive black holes. Here, we present the first ever general relativistic hydrodynamical simulations of the collapse of rapidly accreting supermassive stars. We find that black hole formation is in many cases prevented by nuclear burning due to the long timescales of the collapse of these stars ($10^6$ s). Consequently, this is a novel astrophysical site for hot CNO burning and hydrogen burning via proton captures. For Pop III accreting supermassive stars, we find that only stars with very high (100 $\rm{M_\odot}/$yr) or low (0.1 $\rm{M_\odot}/$yr) accretion rates can form black holes, with models in between undergoing energetic thermonuclear pulsations. The final fate of these pulsating models may be to undergo subsequent pulsations or explosions or to collapse to black holes. For metal rich accreting supermassive stars ($Z\geq 0.1 \rm{Z_\odot}$), we do not find any black hole formation, with some models undergoing extremely energetic explosions ($10^{55}$ ergs). Our results invite further study on the formation of massive black holes from rapidly accreting supermassive stars which have reached the general relativistic radial instability. 

\end{abstract}

\maketitle


\section{\label{sec:intro}Introduction}

The observation of massive black holes in the form of quasars at high redshift \citep{banados2018,wang2021} has puzzled astronomers for decades \citep[e.g.][]{bromm2003}. Recent observations, particularly by JWST NIRSpec have sharpened our understanding of these distant objects \citep{Eilers2023ApJ...950...68E,Marshall2023A&A...678A.191M} and in one case, that of UHZ-1, led to the discovery of a black hole at much higher redshift than was previously known \citep{Goulding2023ApJ...955L..24G}. The existence of these black holes so early on in the history of the universe is perplexing due to their unknown origin \citep{rees1984}. In particular, massive black holes are thought to grow by accretion at well below the Eddington limit, as they undergo a duty cycle whereby episodes of Eddington accretion are terminated by radiative feedback. The crux of the issue is that even if we allowed Eddington accretion to occur unabated in the early universe, then we would still require massive black holes to exist very soon after the big bang. If we instead posited the existence of stellar mass black holes in the early Universe (as one might expect), then super-Eddington or hyper-Eddington accretion would be required to explain the observed quasars. In the astrophysics community, the most popular approach to this problem is to form massive black holes in the early universe directly from primordial gas \citep{inayoshi2020}. This \textit{direct collapse scenario} is something of a misnomer because there is an intermediate state in between primordial gas and black hole, namely a supermassive star. 

Supermassive stars (SMSs) are roughly defined as stars having masses above 10,000 $\msun$. The main difference between these SMSs and more familiar massive stars is that there is thought to be significant accretion ($\dot{M}>10^{-3} \;\msun/$yr) onto the star even after the onset of hydrogen burning. Indeed, this is how such gigantic masses may be accumulated. SMSs are radiation dominated and usually live very close to the Eddington limit, but they are nevertheless thought to undergo various phases of nuclear burning, namely hydrogen, helium, and carbon/oxygen. The caveat is that because of their extremely large mass, these stars may not survive long enough to reach the later evolutionary stages. This is because they may be destabilized by the general relativistic (GR) radial instability \citep{chandrasekhar1964}, a dynamical instability which can trigger in SMSs during the hydrogen or helium burning phases \citep{Nagele2022MNRAS.517.1584N}. If the GR instability occurs in helium burning, then a non-accreting SMS may explode via the explosive alpha process \citep{chen2014,nagele2020,Nagele2022MNRAS.517.1584N} and if the instability occurs in hydrogen burning, then hydrogen burning may also power an explosion \citep{fuller1986,montero2012,Nagele2023MNRAS.523.1629N}. In most cases, however, the SMS will collapse to a black hole, which can then grow by accretion into a supermassive black hole consistent with the properties of the observed quasars. 

In previous work, GR hydrodynamical tests of explosion versus collapse were employed for non-accreting SMSs \citep{montero2012,Nagele2022MNRAS.517.1584N}, but not for accreting SMSs. This is a critical distinction because although most SMSs will not be undergoing accretion at the time of the GR instability, the most massive black hole seeds which may go on to form the most massive quasars in some scenarios \citep{inayoshi2020}, require rapid accretion ($\dot{M}>1 \;\msun/$yr), thus necessitating that the SMS be undergoing accretion at the time of the GR instability \citep{hosokawa2012,hosokawa2013,umeda2016,woods2017,haemmerle2020,Herrington2023MNRAS.521..463H}. In this letter, we perform these very simulations. We find that accreting SMSs take an order of magnitude longer to form black holes in comparison with their non-accreting cousins. This difference allows for significantly more nuclear burning to take place, in most cases preventing black hole formation. This nuclear burning is a combination of hot CNO and other cyclical processes as well as proton captures on light and intermediate isotopes. For metal rich accreting SMSs formed after galaxy mergers \citep{Mayer2010Natur.466.1082M} or by stellar collisions in nuclear star clusters \citep{Gieles2018MNRAS.478.2461G}, we find that none of our models form black holes immediately after the GR instability and that half of them, in fact, explode. 

We will first describe our methods (stellar evolution, GR instability, and hydrodynamics) and our results before contrasting and contextualizing our work in the final section. 

\section{\label{sec:methods}Methods}

To model the SMSs over evolutionary timescales, we use the HOSHI code \citep{takahashi2018} which is a massive star evolution code (Appendix A). In the non-accreting SMS case, HOSHI has been found to agree very well with KEPLER \citep{chen2014,nagele2020}, though unlike some other codes we use the relativistic density when including the first order post Newtonian TOV correction \citep[cf Fig. 2 of][]{Nagele2022MNRAS.517.1584N}. In this paper, we modify HOSHI to include constant accretion onto the star. The assumption of constant accretion may have some shortcomings particularly for low accretion rates \citep{Sakurai2015MNRAS.452..755S}. Recently, several groups have begun using variable accretion rates taken directly from cosmological simulations \citep[e.g.][]{Woods2021ApJ...915..110W}, but due to the exploratory nature of this study we limit ourselves to constant accretion.  
When the star becomes massive enough or compact enough, radial adiabatic pulsations are no longer damped, causing the star to contract on hydrodynamical timescales, where the small destabilizing effects of general relativity play a critical role \citep{chandrasekhar1964}. The timescales for this contraction are much too short to be resolved by stellar evolution codes, thus necessitating an alternative method of searching for the GR radial instability. After every five timesteps of HOSHI, we check the stability of the SMS using a code developed for this purpose in \citet{Nagele2022MNRAS.517.1584N,Nagele2023MNRAS.523.1629N}. We solve the pulsation equation in spherical, linearized, general relativity \citep{chandrasekhar1964} using an iterative method to determine the fundamental mode frequency of the pulsations \citep{Nagele2022MNRAS.517.1584N,Nagele2023MNRAS.523.1629N}. Our method for finding the GR instability succeeds for numerical polytropes and generally agrees with the results of the GR hydrodynamics code.

Finally, once we have found an unstable model, we pass that model to a GR hydrodynamics code \citep{yamada1997,takahashi2018,nagele2020} (Appendix B). The code includes a nuclear network, such that the material is heated y nuclear reactions and primarily cooled by neutrino reactions. The nuclear network is the principal computational cost for this code, so its extent must be chosen carefully. In the simulations presented here, we are interested in hydrogen burning so we include the proton rich side of the proton-neutron plane, making sure to fully resolve rp (rapid proton capture) breakout from the CNO cycle, including 158 isotopes in total. The hydrodynamics code has three possible outcomes, collapse to a black hole, explosion, or a pulsation which leaves behind a hydrostatic remnant. We have run some of the pulsating models with nuclear reactions turned off in order to confirm that these models are unstable and would form black holes without the energy from nuclear burning. For the pulsating case, we use the escape velocity at shock breakout to determine the quantity of ejected mass. As in previous work using this code \citep{Nagele2022MNRAS.517.1584N}, we have performed a numerical resolution study, although we find that our results are only relevant for the pulsation of the Pop III $\dot{M}=\msun/$ yr case which reaches very high temperatures compared to the other models (Table \ref{tab:summary}). For this case, using a smaller network will cause the star to collapse instead of pulsating, as the energy generation will be underestimated.

\section{\label{sec:results}Results}

\begin{figure}
    ~\vspace{-4mm}\\ 
    \centering
    \includegraphics[width=0.5\textwidth]{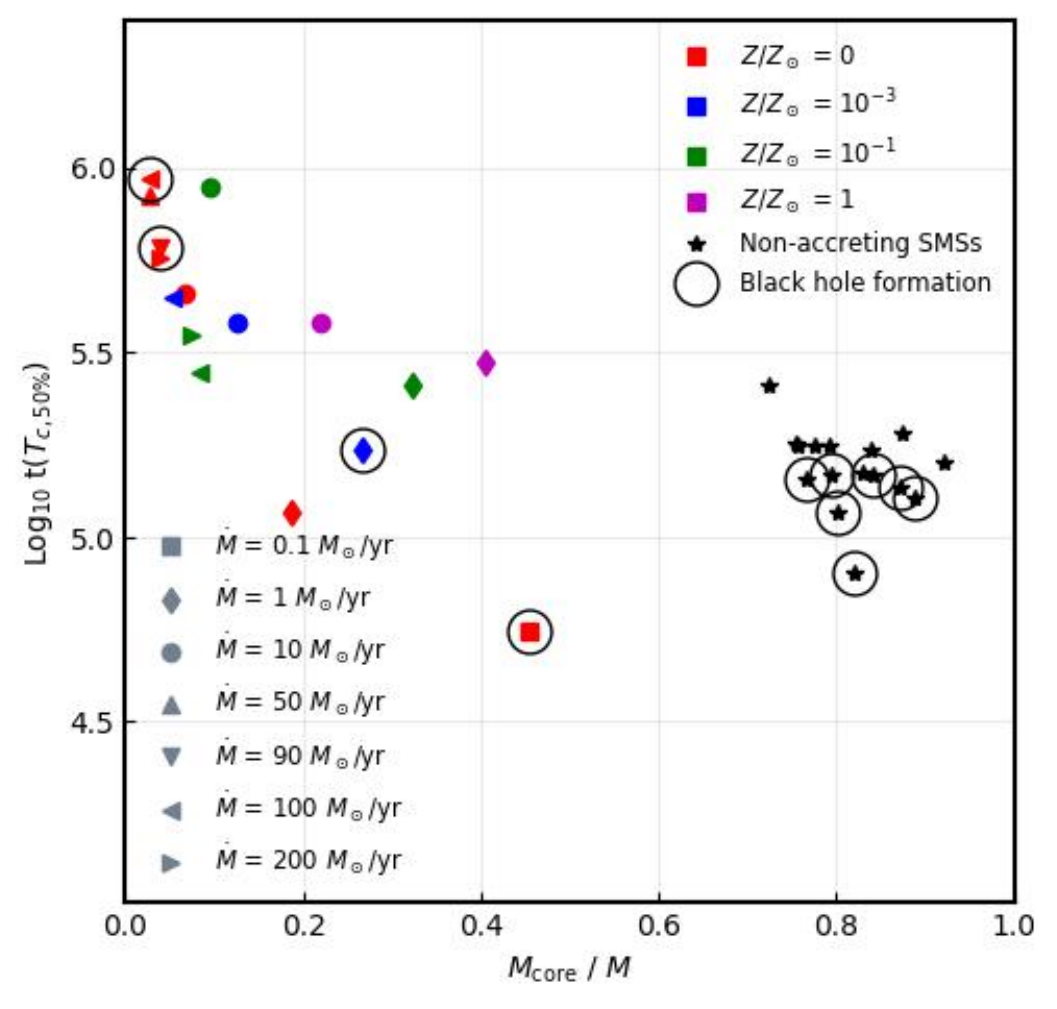}
    \caption{Logarithmic central temperature rise time ($50\%$ greater than the initial value $[$s$]$) after the GR instability as a function of core mass fraction. Colors denote different initial metallicities while circled models go on to form black holes. Black stars are non-accreting SMS models from \citet{Nagele2023MNRAS.523.1629N}. This figure shows three trends: a) black hole formation coincides with the short temperature rise times of more compact models, b) accreting SMSs have longer rise times and c) there is an anti-correlation between core mass fraction and rise times. Taken together, these statements suggest that it is more challenging to form black holes from accreting SMSs than from non-accreting ones, in part because of their small core ratios.}
    \label{fig:mcore_Tc}
\end{figure}

\begin{table*}
\caption{\label{tab:summary}Summary table recording accretion rate, metallicity, central temperature, total mass, and core mass at the GR instability, outcome of the hydrodynamical simulation, maximum temperature, nuclear energy generated, mass ejected and kinetic energy of the ejected material during the hydrodynamical simulation.}
\begin{ruledtabular}
\begin{tabular}{cccccccccc}
 $\dot{M}$ [$\msun$/yr] & Z [$\zsun$] &   Log $T_{c, \rm GRI}$ & $M_{\rm GRI}$ [$\msun$] & $M_{\rm core,GRI}$ [$\msun$] & Outcome  & Log $T_{c, max}$ & $E_{\rm nuc}$ [$10^{54}$ ergs] & $M_{\rm ejected}$  [$\msun$] & $E_{\rm kin}$ [$10^{54}$ ergs]
\\ \hline
0.1 & 0 & 8.184 & 5.325e4 & 2.375e4 & Collapse & --- & ---  & ---  & --- \\
1 & 0 & 8.202 & 9.356e4 & 1.756e4 & Pulsation & 8.709 & 1.059 & 3051 & 0.1312\\
10 & 0 & 8.199 & 1.418e5 & 9349 & Pulsation & 8.471 & 0.6154 & 0 & 0\\
50 & 0 & 8.226 & 2.138e5 & 5881 & Pulsation & 8.519 & 1.138 & 1046 & 0.01036\\
90 & 0 & 8.239 & 2.648e5 & 1.02e4 & Collapse & --- & ---  & ---  & --- \\
100 & 0 & 8.253 & 2.742e5 & 7148 & Collapse & --- & ---  & ---  & --- \\
200 & 0 & 8.254 & 1.846e5 & 6954 & Pulsation & 8.649 & 1.792 & 3878 & 0.1255\\
1 & $10^{-3}$ & 8.012 & 1.305e5 & 3.411e4 & Collapse & --- & ---  & ---  & --- \\
10 & $10^{-3}$ & 8.055 & 1.791e5 & 2.204e4 & Pulsation & 8.343 & 0.88 & 0 & 0\\
100 & $10^{-3}$ & 8.124 & 1.4e5 & 7364 & Pulsation & 8.47 & 0.8506 & 0 & 0\\
1 & $10^{-1}$ & 7.868 & 1.542e5 & 4.886e4 & Pulsation & 8.236 & 1.27 & 4312 & 0.1733\\
10 & $10^{-1}$ & 7.886 & 2.484e5 & 2.25e4 & Pulsation & 8.176 & 0.9747 & 0 & 0\\
100 & $10^{-1}$ & 8.032 & 1.651e5 & 1.35e4 & Explosion & 8.297 & 4.725 & 1.651e5 & 1.043\\
200 & $10^{-1}$ & 8.013 & 1.409e5 & 1.013e4 & Explosion & 8.28 & 3.196 & 1.409e5 & 0.3221\\
1 & 1 & 7.808 & 2.986e5 & 1.188e5 & Explosion & 8.274 & 17.26 & 2.986e5 & 10.3\\
10 & 1 & 7.805 & 2.076e5 & 4.294e4 & Pulsation & 8.069 & 0.8976 & 0 & 0\\
\end{tabular}
\end{ruledtabular}
\end{table*}

\begin{figure}
    ~\vspace{-4mm}\\ 
    \centering
    \includegraphics[width=0.48\textwidth]{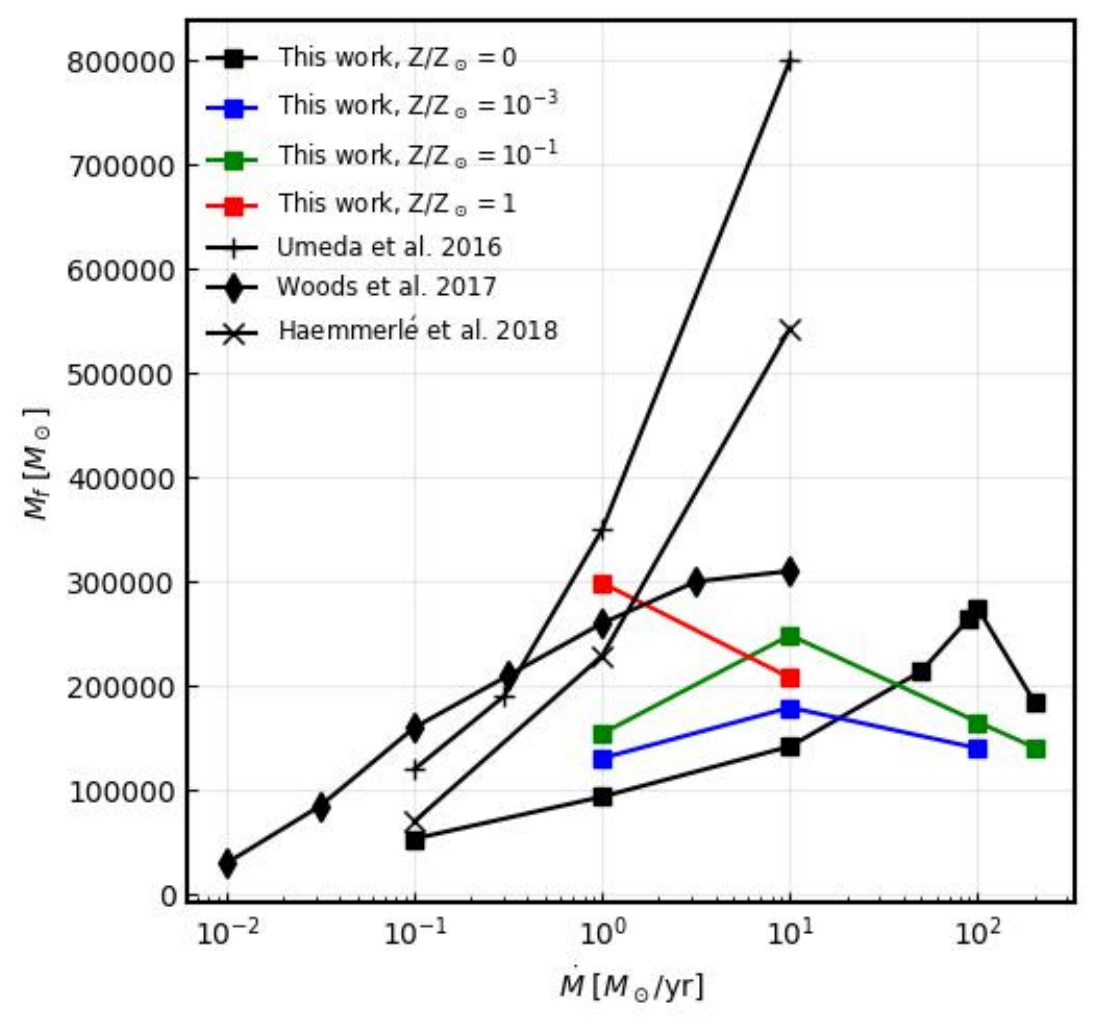}
    \caption{Final mass as a function of mass accretion rates for the four metallicities explored in this paper (Table \ref{tab:summary}) as well as the metal free models from \citet{umeda2016,woods2017,Haemmerle2018MNRAS.474.2757H}. Our metal free models have smaller final masses than previous papers. Also of note is the decreasing final mass at very high accretion rates, not seen in other works.
  }
    \label{fig:mdot_mf}
\end{figure}

\begin{figure}
    ~\vspace{-4mm}\\ 
    \centering
    \includegraphics[width=0.5\textwidth]{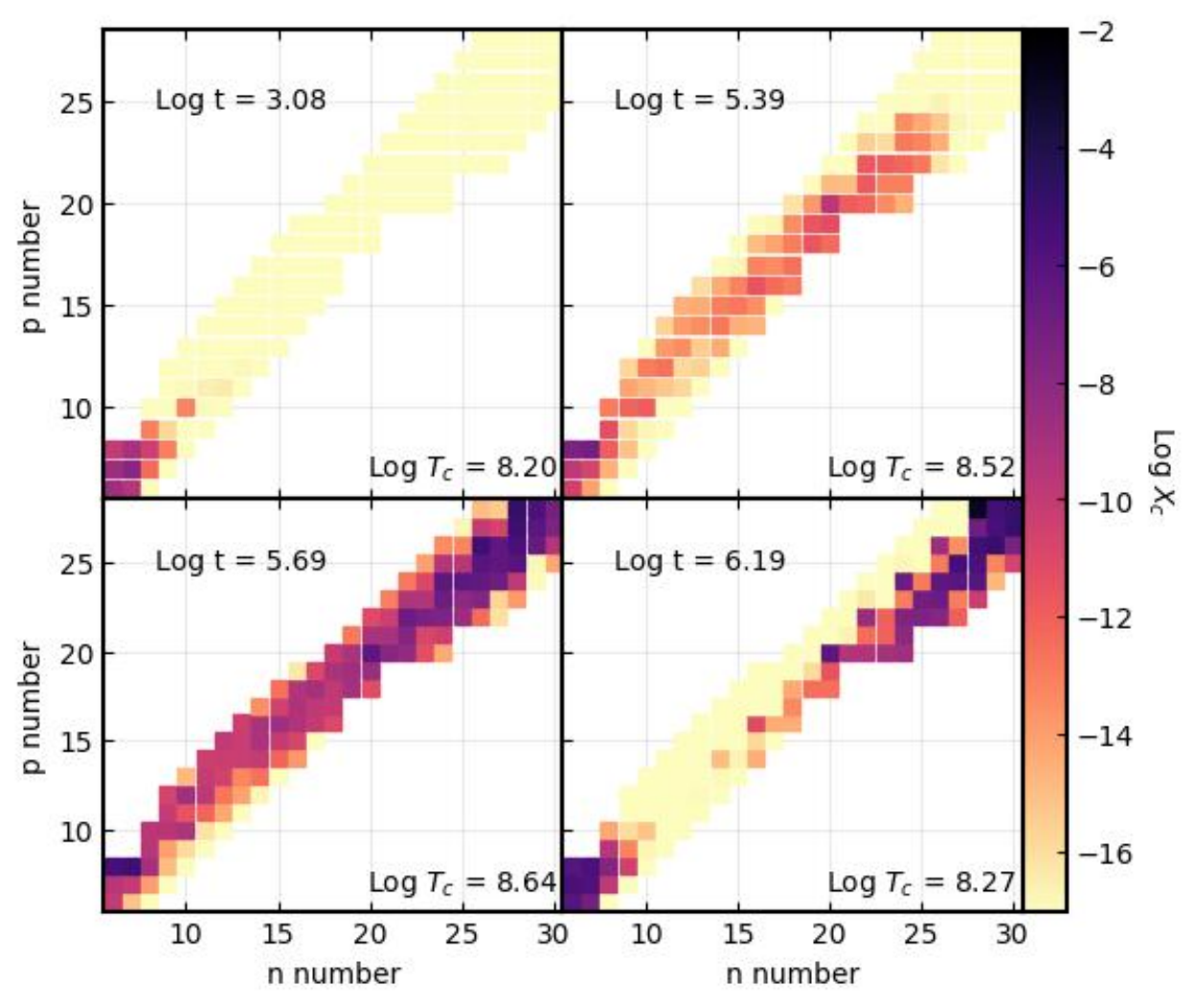}
    \caption{Central isotopic mass fractions for several time snapshots during a pulsation in the Pop III model with an accretion rate of one solar mass per year. The axes are proton number and neutron number of the isotopes while the logarithmic time and temperature are indicated in each panel. The initial composition is entirely CNO and lighter elements. 
  }
    \label{fig:np}
\end{figure}

In this letter, we present seven Pop III models, three with $Z=10^{-3}\zsun$, four with $Z=10^{-1}\zsun$ and two with solar metallicity. Though solar and super-solar metallicities are of particular interest for the galaxy merger scenario \citep{Mayer2010Natur.466.1082M,Mayer2024ApJ...961...76M}, they are much more expensive than the lower metallicity models due to the increased radiative pressure which makes the star more likely to go super-Eddington in the envelope. When such a situation arises, the time step becomes smaller making the total calculation time longer. The most interesting result from the stellar evolution simulations is the core mass to total mass ratio of the accreting SMSs is much lower than in the non-accreting case and decreases with increasing accretion rate (Fig. \ref{fig:mcore_Tc}).

Table \ref{tab:summary} shows the central temperature and mass of each of the models at the GR instability. For the Pop III case, the final masses increase with accretion rate. For instance, $\dot{M} = 0.1\;\msun/$yr has a final mass of just 53,000 $\msun$. The final mass increases up to 274,000 $\msun$ for $\dot{M} = 100\;\msun/$yr, but we note that this increase is not monotonic as some extremely high accretion rates such as $\dot{M} = 200\;\msun/$yr experience the instability at lower masses (Fig \ref{fig:mdot_mf}). This trend is seen across all of our metallicities, and it suggests an upper limit for possible black hole masses of about 300,000 $\msun$ in our models. We note that such behavior is not found by other studies \citep[e.g.][]{haemmerle2020}. The reasons for the eventual decrease in the maximum mass are not immediately clear, and we leave their investigation to future work. We note, however, that these results have been verified with the GR hydrodynamics code, making them slightly more trustworthy than studies which have relied on stability analyses or post Newtonian codes. There is also a metallicity dependence to the final mass, which follows from the increased opacity making metal rich stars less compact and thus less likely to experience the instability. 

We now turn to the hydrodynamics. We will first describe the nucleosynthesis in one extreme example which only barely avoids black hole formation (Pop III, $\dot{M} = 1 \;\msun$/yr). Fig. \ref{fig:np} shows the central isotopic mass fraction for several equally spaced time snapshots. At the beginning of the simulation (first panel), the composition is dominated by primordial elements (hydrogen, helium, lithium, etc) as well as CNO elements which have been synthesized during the stellar evolution. The second panel shows the beginning of hydrogen burning in the lower part of the panel as well as oxygen burning and subsequent proton captures in the middle part of the panel. In the next panel, the central temperature has passed Log $T_c = 8.5$ for more than $10^5$ seconds. We note the long timescale because hydrogen burning via repeated proton captures in this temperature regime is fundamentally limited by inverse beta decay. Thus, for timescales shorter than the ones in this simulation, less burning will take place and it becomes much more likely that a black hole will form. As the temperature continues to increase, more carbon is synthesized at the bottom left of each panel through triple alpha reactions, and that carbon then feeds into the hydrogen burning. This eventually results in quite high mass fractions of nickel (greater than 0.01) in the center of the star. The energy production eventually arrests the collapse and reverses the velocity. the temperature then falls back below it's starting value after which the star stabilizes. This reversal coincides with an outgoing velocity shock which ejects 3051 $\msun$ of material when it reaches the surface of the star. The rest of the SMS then becomes hydrostatic which terminates the simulation. We emphasize that the situation described in this letter is a new astrophysical application for hot CNO burning and proton captures at relatively low temperatures compared to the typical rp process (Fig. \ref{fig:np}).

Fig. \ref{fig:mcore_Tc} shows the time it takes each of the accreting models to increase the central temperature by $50\%$ after the GR instability. For the non-accreting SMSs from \citet{Nagele2023MNRAS.523.1629N}, we see that models with more rapid collapses tend to form black holes. This trend is also present for the accreting SMSs, and since the accreting SMSs take half an order of magnitude longer to collapse, it is more challenging for them to form black holes immediately following the GR instability. 

Four of our models collapse to form black holes: three Pop III models and one $Z=10^{-3} \zsun$ model. We will fist focus on the bottom section of Fig. \ref{fig:mcore_Tc}, which includes the Pop III $\dot{M} = 0.1\;\msun/$yr model. This model most closely resembles the non-accreting SMSs, having a relatively large convective core as well as collapsing on a shorter timescale than the other models. Similar arguments may be applied to the $Z=10^{-3} \zsun$, $\dot{M} = 1\;\msun/$yr model while the nearby Pop III model ($\dot{M} = 1\;\msun/$yr) is the marginal case discussed previously. Thus, all of the metal free and metal poor models in the lower half of the figure either form black holes or come close to doing so. Next, we move to the Pop III $\dot{M} = 90, 100\;\msun/$yr models in the top left. These models collapse slowly after the GR instability, but because their cores are so small, the nuclear burning (which is most efficient in the compact core) may be insufficient to stop the collapse. If this explanation is correct, we must explain why the $\dot{M} = 50, 200\;\msun/$yr models are not able to form black holes, despite having similarly small cores. We note that the $\dot{M} = 90, 100\;\msun/$yr have significantly larger total masses (Table \ref{tab:summary}), and it may be this combination of a large total mass and a small core ratio which permits black hole formation. We stress that the pulsating models leave behind massive remnant SMSs which may themselves go on to experience the GR instability and form black holes, a topic which we plan to explore further in the future.

For the $Z=0.1 \zsun$ and solar metallicity cases. We never find black hole formation. This is because significant seed metals exist enabling hydrogen burning to begin immediately as in \citet{Nagele2023MNRAS.523.1629N}. Indeed, in all of these cases, the maximum temperature is low (Table \ref{tab:summary}) and only a few proton captures on each seed nucleus are required to produce enough energy to stabilize the star. It is thus difficult to see how metal rich accreting stars can form black holes without first transitioning to more compact non-accreting models.

Table \ref{tab:summary} also contains the nuclear energy generated ($E_{\rm nuc}$), the mass ejected by the pulsation or explosion ($M_{\rm ejected}$) and the kinetic energy of the ejected mass ($E_{\rm kin}$, which is the explosion energy for those cases). Unlike in previous work, we find some pulsating models which do not eject significant mass, and we believe the salient difference to be the larger initial mass of the SMS compared with \citet{Nagele2022MNRAS.517.1584N}. The threshold for mass ejection appears to be $10^{54}$ ergs of nuclear energy produced, compared to a typical binding energy of a few times this value.

\section{\label{sec:discussion}Discussion}

Although black hole formation from accreting SMSs has largely thought to have been easier than in the case of non-accreting SMSs, we have shown that this understanding is backwards. Accreting SMSs have very small cores which collapse slowly after the GR instability, thus allowing additional orders of magnitude of nuclear energy generation which can stabilize the star. We will now discuss our results in the context of previous work, astronomical observables, and shortcomings with an eye towards future work. 

There are two sets of previous studies which are relevant to this letter, hydrodynamical simulations of non-accreting SMS explosions \citep{fuller1986,montero2012,chen2014,nagele2020,Nagele2022MNRAS.517.1584N,Nagele2023MNRAS.520L..72N} and evolutionary/hydrodynamical post Newtonian studies of accreting SMSs \citep{woods2017,Herrington2023MNRAS.521..463H}. As we have already contrasted our results with the former, we will now focus on the latter. These studies used stellar evolution codes which become hydrodynamical after some instability is encountered. This approach has previously been shown to find SMS explosions \citep{chen2014}. That being said, there is a large difference in timescales between evolutionary hydrogen burning (thousands of years) and the collapse of the SMS (days to weeks) and there is inherent difficulty in modeling both of these processes using a single code. In our experience with non-accreting SMSs \citep{nagele2020,Nagele2022MNRAS.517.1584N}, this results in the stellar evolution code missing the GR instability and only collapsing at a later time when the star is much more unstable, and thus cannot experience a pulsation or explosion. Another drawback of using the post Newtonian approximation is that even though it very accurately captures the correction to the pressure gradient for a static star, it is missing the post Newtonian dynamical terms which are not easy to incorporate. Finally, our simulations use larger nuclear networks than previous works, and so can better capture energy generation on the dynamical timescale. 

When it comes to observables, the two most promising are the lightcurves of pulsations and explosions \citep{Nagele2023MNRAS.520L..72N} and the nitrogen rich ejecta of the explosions \citep{Nagele2023MNRAS.523.1629N,Nagele2023ApJ...949L..16N}. We note that this nitrogen rich material is only found in the metal rich accreting SMSs which completely explode, as the mass-ejecting pulsations of the Pop III SMSs only contain material with primordial composition. The explosions in this letter will produce super-solar nitrogen abundances such as was seen in GN-z11 \citep{Cameron2023MNRAS.523.3516C} in large quantities, as was discussed in \citet{Nagele2023ApJ...949L..16N}. We also note that the feature of the SMS which made the pulsations studied in \citet{Nagele2023MNRAS.520L..72N} so bright (visible above redshift thirty) is a large progenitor radius, also a feature of the models presented in this letter. We leave detailed calculations of the lightcurves to future work.

We now turn to the limitations of the current study. First of all, the assumption of spherical accretion onto the SMS may be unphysical, given the extreme angular momentum of the accreting material. We note however, that this problem is also present in regular star formation, the difference being that hydrogen burning is not turned on during the accretion phase. The angular momentum question is one of the outstanding topics in the study of accreting SMSs. Relatedly, we have used convection and mass loss prescriptions motivated by massive star evolution and their validity is not guaranteed in the SMS regime. We are interested to see how well our results translate to other accreting SMS models, particularly the lower final mass at very high accretion rates (Table \ref{tab:summary}). Finally, there is the question of what happens to the pulsating models after the pulsation. For the pulsations which do not eject mass, the evolution may continue with only a slight difference in central composition. On the other hand, for pulsating models with large mass ejection, the accretion flow will likely be disrupted, heralding a transition to non-accreting evolution. Finally, the change of composition in the core due to the pulsation will result in increased opacity during the evolution and increased explodability after the next GR instability. We are currently investigating several of these phenomena. 

In conclusion, we have performed the first ever GR hydrodynamical simulations of the collapse of accreting SMSs after the GR instability. We have found that prolonged nuclear burning arrests the collapse and prevents black hole formation in most cases. We plan to further explore the implications of these results to astronomical observables and to determine the final fates of the pulsating models. 

\section*{Data Availability}

The data underlying this article will be shared on reasonable request to the corresponding author.

\section*{Acknowledgements}

This study was supported in part by JSPS KAKENHI Grant Number 23K20864.

\section*{Appendix A: HOSHI}
\label{sec:appendeixA}

The HOSHI (HOngo Stellar Hydrodynamics Investigator) code solves the equations of stellar structure and hydrodynamics in the later stages using a Henyey type implicit
method \citep{takahashi2018}. The code uses the 1st order approximation to the TOV equation in order to include the static correction to Newtonian gravity \citep{nagele2020}, where we absorb the isotopic mass excess into the internal energy density \citep{Nagele2022MNRAS.517.1584N}. The code includes a 52 isotope nuclear network \citep{Cyburt2010ApJS..189..240C}, neutrino cooling \citep{itoh1996}, an analytical equation of state \citep{Blinnikov1996ApJS..106..171B,takahashi2016}, and convective and radiative energy transfer using the OPAL Project's Rosseland mean opacities \citep{Iglesias1996ApJ...464..943I}.

In this paper, we have updated the code to include constant accretion onto the star, following established methods \citep{Hosokawa2009ApJ...691..823H,hosokawa2012,hosokawa2013,Paxton2015ApJS..220...15P,umeda2016,woods2017,Haemmerle2018MNRAS.474.2757H,Herrington2023MNRAS.521..463H}. We initialize the models in this paper as 1000 $\msun$ pre-main sequence stellar evolution models, where by pre-main sequence we mean that the temperature and density are low. These models thus contract rapidly after the beginning of the calculation, reaching quasi-stable states at masses of between $1100$ $\msun$ and $1500$ $\msun$ for the accretion rates considered in this paper. This rapid contraction is not physically motivated, and we thus do not trust our results below about $2000$ $\msun$, but since the mass ranges of interest in this paper are in excess of $10^5$ $\msun$, this approximation is reasonable. In order to include the accreting mass in subsequent time steps, we increase the mass in the outer $1\%$ of the star by the appropriate amount. The new material is assumed to have the same composition (the metallicity as the star is never fully convective) and entropy \citep{hosokawa2013,umeda2016} as the material in the outer part of the accretion envelope. Because this material is heavier than the corresponding meshes in the previous timestep, compressional heating transports heat away from the outer boundary. In previous studies using HOSHI, rezoning of the mesh was carried out whenever jumps in temperature or density emerged in the stellar interior, increasing the resolution to a maximum of between 1200 and 1500 meshpoints. For the accretion rates in this paper this procedure does not require alteration. For modeling even larger accretion rates, we do have the ability to increase the meshpoint number further if necessary.

\section*{Appendix B: Hydrodynamics}
\label{sec:appendeixB}

In this paper, we simulate the hydrodynamics using a 1D GR Lagrangian hydrodynamics code originally developed for core collapse supernovae \citep{yamada1997,sumiyoshi2005} which has been modified for thermonuclear supernovae with the coupling of a nuclear network and the removal of neutrino transfer \citep{takahashi2018}, and specifically adapted to the explosions of supermassive stars \citep{nagele2020,Nagele2022MNRAS.517.1584N,Nagele2023MNRAS.523.1629N}. The code uses a Roe-type approximate linearized Riemann solver to update the timestep implicitly after solving for fifteen independent variables \citep{yamada1997}. The mapping from HOSHI to the hydrodynamics code is done through logarithmic interpolation of the variables onto a mesh with constant mass difference except for the innermost meshpoint. After the simulation is initiated, the timestep is controlled by the Courant condition and the rate of change of the independent variables at each meshpoint. 

One of the key parameters for this code is the size and composition of the nuclear network. In this paper, we use a 158 isotope network (Fig. \ref{fig:np}) which resolves p-side isotopes so that the network can resolve rp breakout and cyclical processes such as hot CNO and NeNa/MgAl. The network extends up to $^{59}$Ni and thus should include all energy producing reactions relevant to this scenario. This large network, however, is only crucial for the highest temperature models, with other models sufficiently resolved by networks terminating before the iron peak.

\bibliography{bib}

\end{document}